\documentclass[conference]{IEEEtran}

\usepackage[cmex10]{amsmath}
\usepackage{amssymb}

\usepackage{cases}
\usepackage{multirow}
\usepackage{empheq}

\usepackage{cite}
\usepackage{verbatim}

\ifCLASSINFOpdf
  \usepackage[pdftex]{graphicx}
\else
  \usepackage[dvips]{graphicx}
\fi
\usepackage[caption=false,font=footnotesize]{subfig}

\usepackage{color}
\usepackage{multirow}

\begin{document}

\sloppy

\title{Coding Scheme for 3D Vertical Flash Memory}

%\author{
%\IEEEauthorblockN{Yongjune Kim and B. V. K. Vijaya Kumar}
%\IEEEauthorblockA{Data Storage Systems Center (DSSC)\\
%    Carnegie Mellon University\\
%    Pittsburgh, PA, USA\\
%    Email: yongjunekim@cmu.edu, kumar@ece.cmu.edu}
%\and
%\IEEEauthorblockN{Robert Mateescu}
%\IEEEauthorblockA{HGST Research\\
%San Jose, CA, USA\\
%Email: robert.mateescu@hgst.com}
%}

\author{\IEEEauthorblockN{Yongjune Kim\IEEEauthorrefmark{1},
Robert Mateescu\IEEEauthorrefmark{2},
Seung-Hwan Song\IEEEauthorrefmark{2},
Zvonimir Bandic\IEEEauthorrefmark{2},
and
B. V. K. Vijaya Kumar\IEEEauthorrefmark{1}}
\IEEEauthorblockA{\IEEEauthorrefmark{1}Data Storage Systems Center (DSSC), Carnegie Mellon University, Pittsburgh, PA, USA\\ Email: yongjunekim@cmu.edu, kumar@ece.cmu.edu}
\IEEEauthorblockA{\IEEEauthorrefmark{2}HGST Research, San Jose, CA, USA\\
Email: \{robert.mateescu, seung-hwan.song, zvonimir.bandic\}@hgst.com}
}

%% To balance the two columns, you should reduce the text-height of
%% the last page using the following command:
%%%%%%%%%%%%%%%%%%%%%%%%%%%%%%%%%%%%%%%%%%%%%%%%%%%%%%%%%%%%%%%%%%%%%
%\addtolength{\textheight}{-9.35cm}
%%%%%%%%%%%%%%%%%%%%%%%%%%%%%%%%%%%%%%%%%%%%%%%%%%%%%%%%%%%%%%%%%%%%%
%% with an appropriate value. This command must be place on the second
%% last page, i.e., for a one-page abstract here, for a two-page
%% abstract right after the \maketitle command.

%% Create the title:
\maketitle

\begin{abstract}
Recently introduced 3D vertical flash memory is expected to be a disruptive technology since it overcomes scaling challenges of conventional 2D planar flash memory by stacking up cells in the vertical direction. However, 3D vertical flash memory suffers from a new problem known as \emph{fast detrapping}, which is a rapid charge loss problem. In this paper, we propose a scheme to compensate the effect of fast detrapping by \emph{intentional} inter-cell interference (ICI). In order to properly control the intentional ICI, our scheme relies on a coding technique that incorporates the side information of fast detrapping during the encoding stage. This technique is closely connected to the well-known problem of \emph{coding in a memory with defective cells}. Numerical results show that the proposed scheme can effectively address the problem of fast detrapping.
\end{abstract}

\section{Introduction}

Aggressive scaling down of the device dimension has driven the continuous growth of flash memory density. However, the scaling down leads to many challenges such as photolithography limitation, increased inter-cell interference (ICI), and disturbance~\cite{Prall2007, Park2014scaling}. In order to overcome these scaling challenges, a paradigm shift from the 2-dimensional (2D) planar structure to the 3D vertical structure is underway. The change is stacking up cells in the vertical direction instead of shrinking cells within a 2D plane~\cite{Tanaka2007, Jang2009, Park2014scaling, Park2014isscc}.

The recent 3D vertical NAND flash memory with 24 word-line (WL) shows better device characteristics compared to 2D 1x nm planar flash memory~\cite{Park2014isscc}. However, 3D vertical flash memory has a problem of \emph{fast detrapping}, which is a quick charge loss phenomenon resulting in larger threshold voltage variations in programmed cells~\cite{Park2014isscc, Chen2010}.

The fast detrapping usually occurs in charge trap cells rather than floating gate cells~\cite{Chen2010, Park2014isscc}. Since 3D vertical flash memory uses charge trap cells such as damascened metal-gate silicon-oxide-nitride-oxide-silicon (SONOS) cells for easier 3D integration~\cite{Tanaka2007, Jang2009, Park2014isscc}, fast detrapping is an important problem of 3D vertical flash memory. On the other hand, the fast detrapping does not happen in 2D planar flash memory consisting of floating gate cells.

In order to cope with fast detrapping, several approaches have been proposed. These approaches include cell structure engineering at device level and reprogramming at circuit level~\cite{Chen2010, Park2014isscc}. In this paper, we try to combat the fast detrapping at coding level. The basic idea is to compensate the charge loss due to fast detrapping through \emph{intentional} ICI. Although the ICI is a well-known adverse effect~\cite{Lee2002}, it can be utilized to alleviate the effect of the fast detrapping through the proposed scheme. It is worth mentioning that the proposed scheme at coding level can cooperate with other approaches at different levels.

We will formulate the problem of controlling the intentional ICI into \emph{coding in a memory with defective cells} model of \cite{Kuznetsov1974}, which is a notable example of coding with side information available at the encoder~\cite{ElGamal2011}. With this formulation, the additive encoding schemes in~\cite{Tsybakov1975additive, Heegard1983plbc} can be applied to control the intentional ICI.

Recently, coding schemes with the side information available at the encoder have drawn attention for phase change memories (PCM), write once memories (WOM), and flash memories~\cite{Jagmohan2010coding, Kurkoski2013, Kim2014dirtyflash}. In~\cite{Jagmohan2010coding}, the side information corresponding to the variability of PCM is used at the encoder to reduce the effect of variability. \cite{Kurkoski2013} assumes that the current state of the memory is known to the encoder for rewriting WOM. Also, the encoder exploits the side information corresponding to the ICI of flash memory and reduces the effect of the ICI in~\cite{Kim2014dirtyflash}. In this paper, the side information is related to the fast detrapping of 3D vertical flash memory, and coding with this side information controls the intentional ICI to compensate the fast detrapping.

Although the above coding schemes improve the reliability (i.e., decoding failure probability) by using the side information at the encoder, the write speed performance would be degraded during the obtaining of the side information and incorporating this side information into encoding. Note that soft decision decoding schemes such as low-density parity check (LDPC) codes in flash memory improve the decoding failure probability at the expense of decreased read speed due to multiple read operations needed to obtain soft decision values. Thus, we can claim that the coding schemes using side information at the encoder are complementary to the soft decision decoding schemes.

In memory systems, it is well-known that the read speed performance is more critical than the write speed performance since the write operation is typically not on the critical path because of write buffers in the memory hierarchy and the write latency can be hidden~\cite{Jagmohan2010coding, Hennessy2002}. Also, the read operations are required more often than the write operations in many memory applications. Thus, the coding schemes using side information at the encoder have an advantage over soft decision decoding from the perspective of speed performance.

%In this paper, we will focus on single-level cell (SLC) 3D vertical flash memories, the extension of proposed scheme to multi-level cell (MLC) flash memories will be a future research topic.

The rest of this paper is organized as follows. Section~\ref{sec:background} explains the basics of 3D vertical flash memory and the channel model of 3D vertical flash memory by taking into account the fast detrapping and the ICI. Section~\ref{sec:proposed} proposes a coding scheme that overcomes fast detrapping by controlling the intentional ICI. After showing the simulation results in Section~\ref{sec:result}, we will conclude the paper in Section~\ref{sec:conclusion}.

\section{Background and Channel Model}\label{sec:background}

\subsection{Basic Operations} \label{subsec:operation}

\begin{figure}[!t]
\centering
\subfloat[Single-level cell (SLC) for $B=1$]{\includegraphics[width=3in]{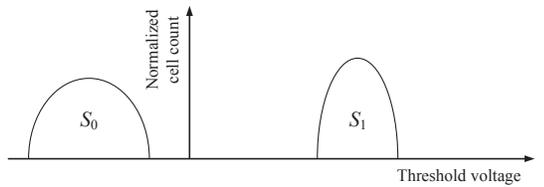}
\label{fig:slc}}
\hfil
\vspace{-1mm}
\subfloat[Multi-level cell (MLC) for $B=2$]{\includegraphics[width=3in]{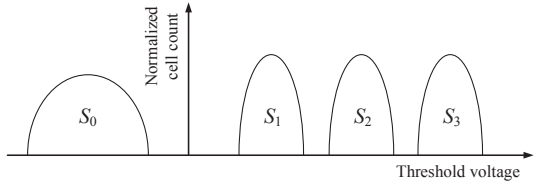}
\label{fig:mlc}}
\caption{Threshold voltage distribution of flash memory cells.}
\label{fig:slc_mlc}
\vspace{-5mm}
\end{figure}

In order to store $B$-bit per cell, each cell's threshold voltage is divided into $2^B$ states, which is similar to pulse amplitude modulation (PAM). Fig.~\ref{fig:slc_mlc}~\subref{fig:slc} shows the threshold voltage distribution of 1-bit per cell flash memory, which is traditionally called single-level cell (SLC). Initially, all memory cells are erased, so their threshold voltage is in the lowest state $S_0$ (erase state). In order to store data, some of the cells in $S_0$ should be written (programmed) into $S_1$. For multi-level cell (MLC) flash memories (i.e., $B \ge 2$), some of cells in $S_0$ will be programmed into $S_1, \ldots, S_{2^B - 1}$ (program states) as shown in Fig.~\ref{fig:slc_mlc}~\subref{fig:mlc}.

%In write operation, the page buffer in Fig.~\ref{fig:structure} is loaded with a unit of page data. Depending on the loaded data in the page buffer, some of cells remain in erase state and others are programmed into program states.

The most widely used write operation scheme is the incremental step pulse programming (ISPP) scheme, which was proposed to maintain a tight threshold voltage distribution for high reliability \cite{Suh1995}. The ISPP is based on repeated program and verify cycles with the staircase program voltage $V_{\textrm{pp}}$. Each program state associates with a verify level that is used in the verify operation. During each program and verify cycle, the cell's threshold voltage is boosted by up to the incremental step voltage $\Delta V_{\textrm{pp}}$ and then compared with the corresponding verify level. If the threshold voltage of the memory cell is still lower than the verify level, the program and verify iteration continues. Otherwise, further programming of this cell is disabled \cite{Suh1995, Dong2011soft}.

The positions of program states are determined by verify levels and the tightness of each program state depends on the incremental step voltage $\Delta V_{\textrm{pp}}$. By reducing $\Delta V_{\textrm{pp}}$, the threshold voltage distribution can be made tighter, however the write time increases \cite{Suh1995, Kim2012verify}.

In read operation, the threshold voltages of cells in the same WL are compared to a given read level. After a read operation, a page of binary data is transferred to the page buffer. The binary data shows whether the threshold voltage of each cell is lower or higher than the given read level. Namely, the read operation of flash memory is a binary decision. Conventionally, if a cell's threshold voltage is lower than the given read level, ``1'' is transferred to the corresponding page buffer. Otherwise, ``0'' is transferred to the page buffer.

Hence, multiple read operations are required to obtain a soft decision value, which lowers the read speed. The degradation of read speed is an important challenge for soft decision decoding in flash memories \cite{Dong2011soft}.

The threshold voltage of flash memory cell can be reduced by erase operation. In flash memory, all the memory cells in the same flash memory block should be erased at the same time \cite{Suh1995}. In addition, the threshold voltage of cell should be moved into the lowest state $S_0$ by erase operation whereas a slight increase of threshold voltage is possible by ISPP during write operation \cite{Suh1995}. %These unique properties of flash memory cause \emph{asymmetry between write and erase operations}.

\subsection{3D Vertical Flash Memory and ICI} \label{subsec:vertical}

\begin{figure}[!t]
\centering
\subfloat[Simplified bird's eye view of vertical channel type 3D flash memory cell array]{\includegraphics[width=3.2in]{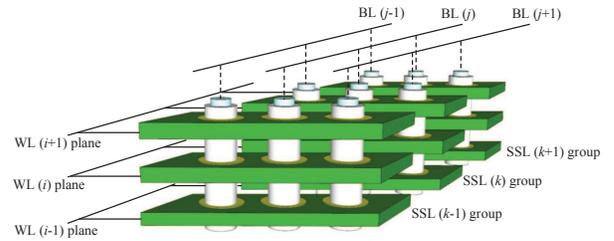}
\label{fig:vertical_3d}}
\hfil
\vspace{-0.5mm}
\subfloat[Cross section diagram along single WL plane]{\includegraphics[width=2.2in]{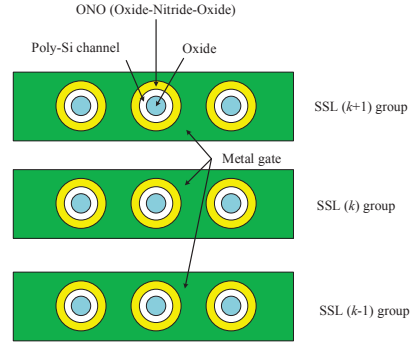}
\label{fig:vertical_2d}}
\caption{3D vertical flash memory cell array in~\cite{Park2014isscc}.}
\label{fig:vertical}
\vspace{-5mm}
\end{figure}

Among various 3D flash memory array architectures, the vertical channel type architecture having multiple WL planes, i.e., 3D vertical flash memory has been adopted in~\cite{Tanaka2007, Jang2009, Park2014isscc} for easier 3D integration and better device characteristics. Fig.~\ref{fig:vertical} illustrates the simplified bird's eye view of 3D vertical flash memory cell array in~\cite{Park2014isscc} and its cross section diagram along single WL plane. Note that a string-select-line (SSL) group in Fig.~\ref{fig:vertical}~\subref{fig:vertical_3d} is equivalent to the 2D planar flash memory cell array.

In contrast to the 2D planar flash memory where the floating gate is formed as an electron storage layer upon single crystal planar silicon channel, the 3D vertical flash memory has a nitride layer inside oxide-nitride-oxide (ONO) stack which is grown as a charge trap layer along the circumference of the thin poly-silicon vertical channel. Note that each charge trap layer in this 3D vertical flash memory is surrounded by the metal gates along WL plane.

By taking into account the structure of 3D vertical flash memories, we will address the ICI of 3D vertical flash memories. In flash memory, the threshold voltage shift of one cell affects the threshold voltage of its adjacent cell because of the ICI. The ICI is mainly attributed to parasitic capacitances coupling effect between adjacent cells~\cite{Lee2002}.

Suppose that $V_{(i, j, k)}$ is the threshold voltage of $(i, j, k)$ cell which is situated at $i$-th WL, $j$-th bit-line (BL), and $k$-th SSL as shown in Fig.~\ref{fig:vertical}. The threshold voltage shift $\Delta_{\textrm{ICI}}V_{(i, j, k)}$ of the $(i,j,k)$ cell due to the ICI can be given by
\begin{equation} \label{eq:ici}
\begin{aligned}
\Delta_{\textrm{ICI}}V_{(i, j, k)} &= \gamma_{\textrm{WL-to-WL}} \left( \Delta V_{(i-1, j, k)} + \Delta V_{(i+1, j, k)} \right) \\
    &+ \gamma_{\textrm{BL-to-BL}} \left( \Delta V_{(i, j-1, k)} + \Delta V_{(i, j+1, k)} \right) \\
    &+ \gamma_{\textrm{SSL-to-SSL}} \left( \Delta V_{(i, j, k-1)} + \Delta V_{(i, j, k+1)} \right)\\
\end{aligned}
\end{equation}
which is an extension of the ICI model of 2D planar flash memories in~\cite{Lee2002, Kim2013modulation}. $\Delta V_{(i\pm1, j\pm1, k\pm1)}$ in the right hand side represent the threshold voltage shifts of adjacent cells after the $(i, j, k)$ cell has been written. Note that $\gamma_{\textrm{WL-to-WL}}$ is coupling ratio between WL plane and adjacent WL plane. Also, $\gamma_{\textrm{BL-to-BL}}$ is coupling ratio between BL and adjacent BL. Finally, $\gamma_{\textrm{SSL-to-SSL}}$ is coupling ratio between SSL group and its adjacent SSL group. It is worth mentioning that the diagonal ICIs are neglected since they are very small due to the longer distance between cells.

Since each charge trap layer in this 3D vertical flash memory is surrounded by the metal gates along WL plane, the ICI between adjacent BLs typically found in high density 2D planar flash memory is completely absent in the same WL plane. Similarly, the ICI between adjacent SSL groups will be negligible due to the metal gates.

The ICI between adjacent WL planes is also reduced compared to the conventional 2D planar flash memory since the charge trap layer in the 3D vertical flash memory is much thinner than the floating gate layer in the 2D flash memory. However, the ICI between adjacent WL planes increases as the distance between WL planes is reduced for the higher cell density. Thus, it is enough to take into account only the ICI between adjacent WL planes in the same SSL group and BL. By setting $\gamma = \gamma_{\textrm{WL-to-WL}}$ and $\gamma_{\textrm{BL-to-BL}} = \gamma_{\textrm{SSL-to-SSL}} = 0$, the ICI model of \eqref{eq:ici} will be simplified into
\begin{equation} \label{eq:ici_simple}
\Delta_{\textrm{ICI}}V_{(i, j, k)} = \gamma \left( \Delta V_{(i-1, j, k)} + \Delta V_{(i+1, j, k)} \right).
\end{equation}

\subsection{Fast Detrapping} \label{subsec:detrapping}

As mentioned in~\ref{subsec:vertical}, nitride charge trap layer is attractive for integration of the 3D vertical flash memory. However, the fast detrapping from this trap layer is a critical challenge in the 3D vertical flash memory~\cite{Chen2010}.

\begin{figure}[!t]
\centering
\subfloat[Threshold voltage distribution of cells in the $i$-th WL before fast detrapping]{\includegraphics[width=2.2in]{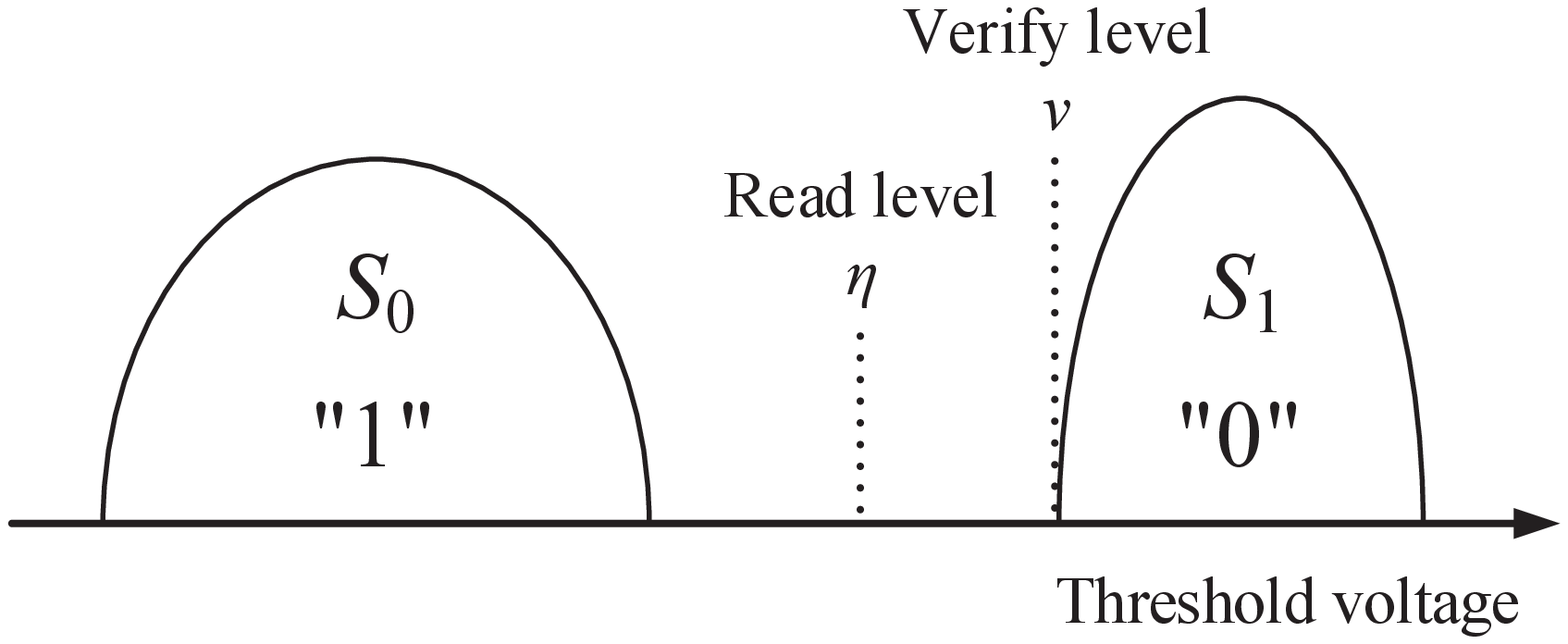}
\label{fig:detrapping_before}}
\vspace{-3mm}
\hfil
\subfloat[Threshold voltage distribution of cells in the $i$-th WL after fast detrapping]{\includegraphics[width=2.2in]{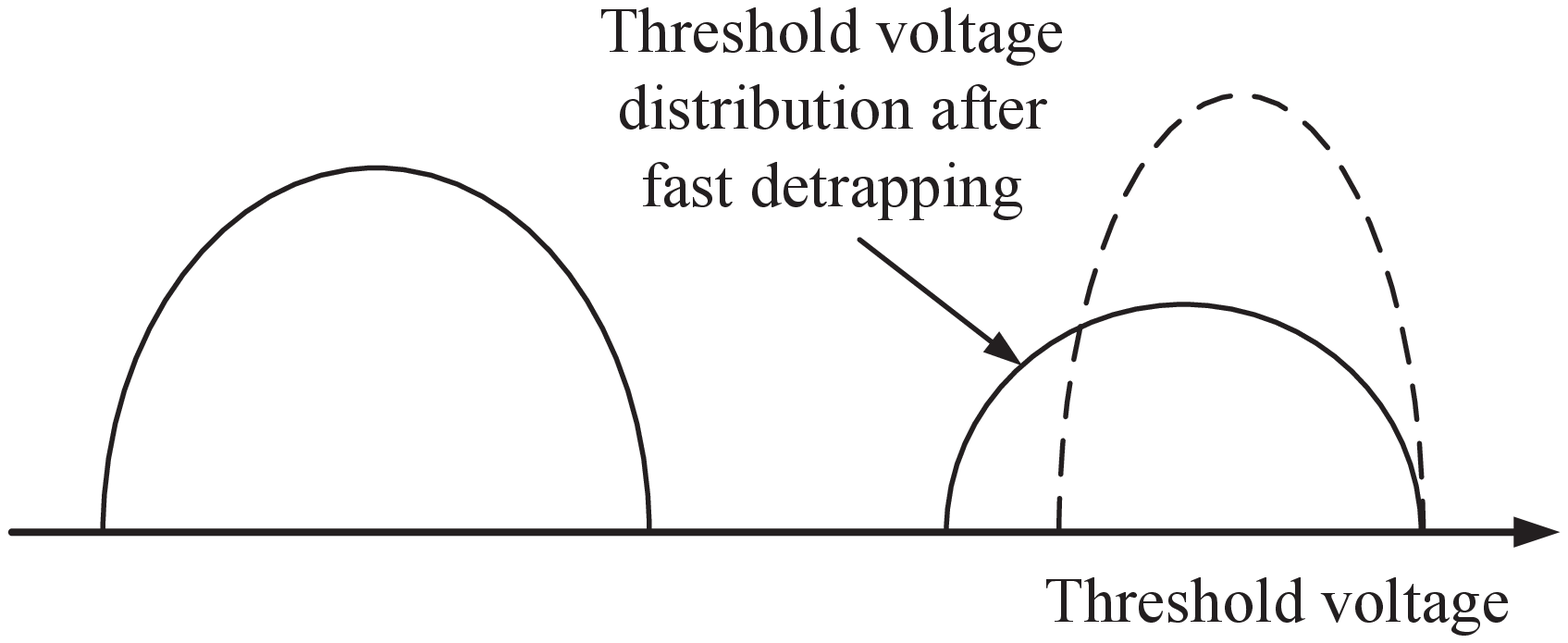}
\label{fig:detrapping_after}}
\vspace{-3mm}
\hfil
\subfloat[Identified cells that suffer from fast detrapping in the $i$-th WL]{\includegraphics[width=2.2in]{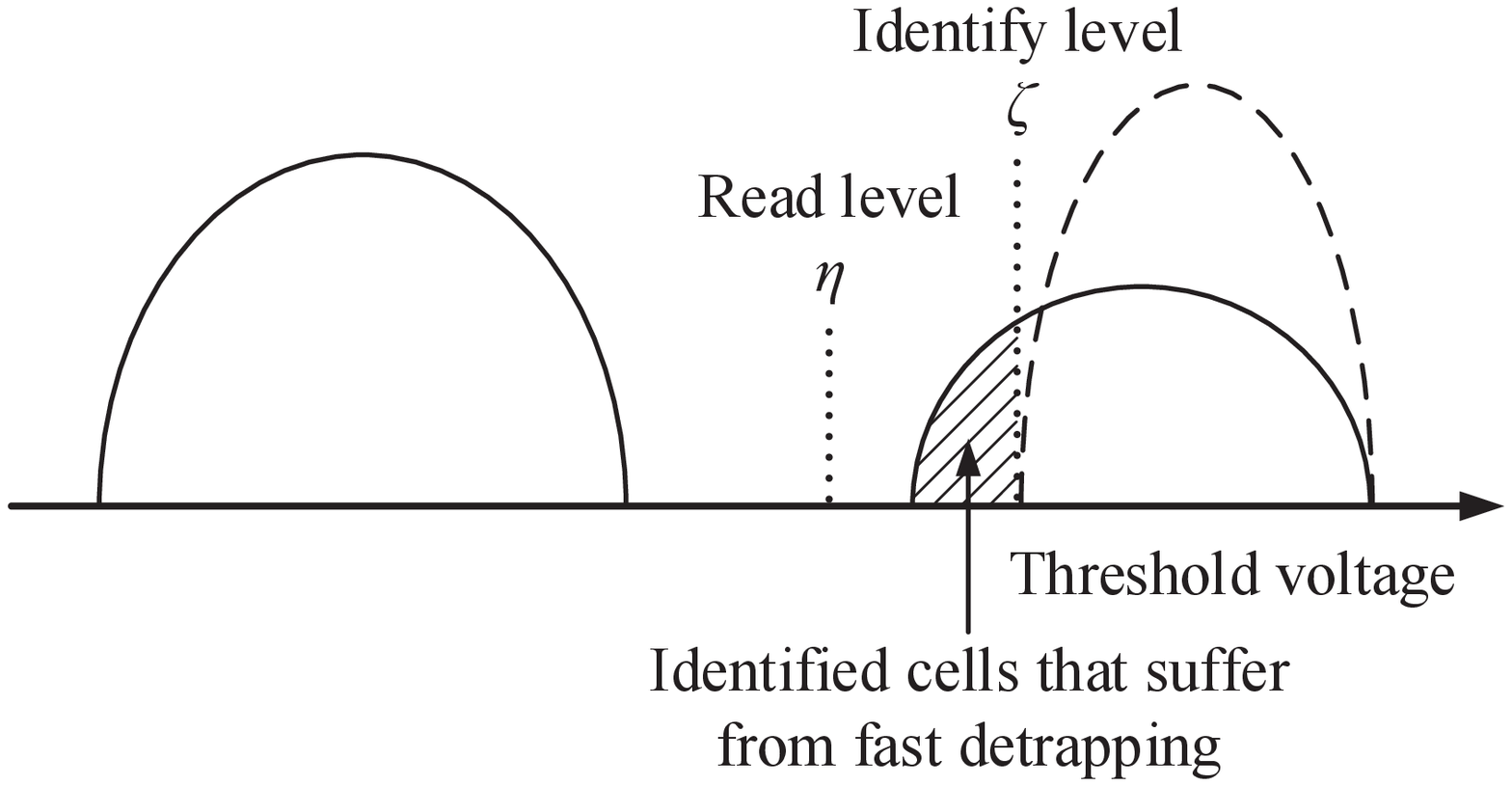}
\label{fig:detrapping_identify}}
\caption{Fast detrapping and identifying cells that suffer from fast detrapping.}
\label{fig:detrapping_before_after}
\vspace{-5mm}
\end{figure}

The cause of fast detrapping can be explained by shallowly trapped electrons in charge trap layers, which immediately detrap and tunnel out after the programming pulse is terminated~\cite{Chen2010}. Thus, the charge loss due to fast detrapping quickly decreases the threshold voltage of corresponding cells and degrades the threshold voltage distribution, as shown in Fig.~\ref{fig:detrapping_before_after}~\subref{fig:detrapping_before} and~\subref{fig:detrapping_after}. The fast detrapping occurs immediately after write operation~\cite{Chen2010}.

The device level approaches such as cell structure and material could not completely solve this problem~\cite{Chen2010}. In~\cite{Park2014isscc}, a counter-pulse program using self-boosting was proposed at circuit level, which accelerates fast detrapping before a verify operation of ISPP such that fast detrapped cells can be reprogrammed by the subsequent programming pulses. Nevertheless, fast detrapping can happen again in the later programming pulses, which requires a different approach at higher levels such as coding level. Since the approaches at different levels can coexist, the proposed scheme at coding level can cooperate with other approaches at lower levels in combating fast detrapping.

%In this work, we propose a new coding scheme to help solving this fast de-trapping issue. This technique is also generally applicable as compensation technique of charge loss issues which is especially highly concerned in Triple-Level Cell (TLC) NAND flash memories requiring tight cell threshold voltage distributions.

\subsection{Channel Model} \label{subsec:channel_model}

A channel model of conventional 2D planar flash memories can be given by
\begin{align}
Y &= X + S_{\textrm{2D}} + Z \label{eq:fmc0} \\
  &= X + Z_\textrm{write} + S_{\textrm{2D}} + Z_\textrm{read} \label{eq:fmc1} \\
  &= V + S_{\textrm{2D}} + Z_\textrm{read} \label{eq:fmc2}
\end{align}
where $X$ and $Y$ are the channel input and output. Also, $S_{\textrm{2D}}$ represents the ICI from adjacent cells in the conventional 2D planar flash memory. The additive random noise $Z$ is a sum of $Z_\textrm{write}$ and $Z_\textrm{read}$ where $Z_\textrm{write}$ is the write noise due to the initial threshold voltage distribution after erase operation and the incremental step voltage $\Delta V_{\textrm{pp}}$ of ISPP. $Z_\textrm{read}$ is the read noise due to other noise sources.

Since the write noise $Z_\textrm{write}$ precedes the ICI $S_{\textrm{2D}}$, we consider a random variable $V = X + Z_\textrm{write}$. The shifts of $V$ in adjacent cells determine the ICI $S_{\textrm{2D}}$. \eqref{eq:fmc0} and \eqref{eq:fmc1} come from the flash memory channel model in \cite{Moon2013, Kim2014dirtyflash}. The channel model of \eqref{eq:fmc0} was recently validated by experimental results from the 2x nm 2D planar flash memory in~\cite{Moon2013}.

The 3D vertical flash memory channel model can be extended from the 2D planar flash memory channel model as follows.
\begin{align}
Y &= V + S_\textrm{3D} + Z_\textrm{read} \label{eq:fmc3} \\
  &= V + S_\textrm{3D} + Z_\textrm{fast} + Z_\textrm{random} \label{eq:fmc4}
\end{align}
where $Z_\textrm{read} = Z_\textrm{fast} + Z_\textrm{random}$. $Z_\textrm{fast}$ denotes the noise due to fast detrapping. All the other read noise sources are represented by $Z_\textrm{random}$. In addition, the ICI of the 3D vertical flash memory is given by $S_\textrm{3D}$. Note that $S_\textrm{3D}$ of $(i, j, k)$ cell is the same as $\Delta_{\textrm{ICI}}V_{(i, j, k)}$ of \eqref{eq:ici} and \eqref{eq:ici_simple}.

\section{Proposed Scheme}\label{sec:proposed}

We propose a new scheme at coding level in order to overcome fast detrapping. Since the proposed scheme will be explained by the problem of \emph{coding with side information available at the encoder}, we should address the following two questions.

\begin{itemize}
  \item How does the encoder obtain the side information?
  \item How does the encoder use the side information?
\end{itemize}

We will address these two problems and propose the solutions in the following subsections.

\subsection{How Does the Encoder Obtain the Side Information?} \label{subsec:know}

First, the encoder should obtain the side information corresponding to fast detrapping. Fig.~\ref{fig:detrapping_before_after}~\subref{fig:detrapping_before} shows the threshold voltage distribution before fast detrapping happens. After writing the $i$-th WL, some of cells in the $i$-th WL immediately suffer from fast detrapping and their threshold voltages decrease as shown in Fig.~\ref{fig:detrapping_before_after}~\subref{fig:detrapping_after}.

The cells suffering from the fast detrapping can be identified by two read operations at the read level $\eta$ and the identify level $\zeta$, as shown in Fig.~\ref{fig:detrapping_before_after}~\subref{fig:detrapping_identify}. If a cell's threshold voltage is between $\eta$ and $\zeta$, we can claim that this cell suffers from fast detrapping and the encoder obtains the location of this cell, which is the side information about fast detrapping.

If the encoder holds the original data of the $i$-th WL, we can obtain this side information of fast detrapping by just one read operation since we do not need to apply the read operation at the read level $\eta$. By combining the original data and the binary data obtained from the read operation at the identify level $\zeta$, the encoder can identify the cells suffering from fast detrapping.

Note that the identify level $\zeta$ do not need to be the same as the verify level $\nu$. We can change the identify level $\zeta$ by taking into account the strength of fast detrapping. The number of identified cells depend on the identify level $\zeta$.

\subsection{How Does the Encoder Use the Side Information?} \label{subsec:use}

Now the encoder knows the side information corresponding to fast detrapping in cells of the $i$-th WL. The side information represents the locations of identified cells, which suffer from fast detrapping. The next step is to use this side information during the encoding stage.

\begin{figure}[t]
   \centering
   \includegraphics[width=0.45\textwidth]{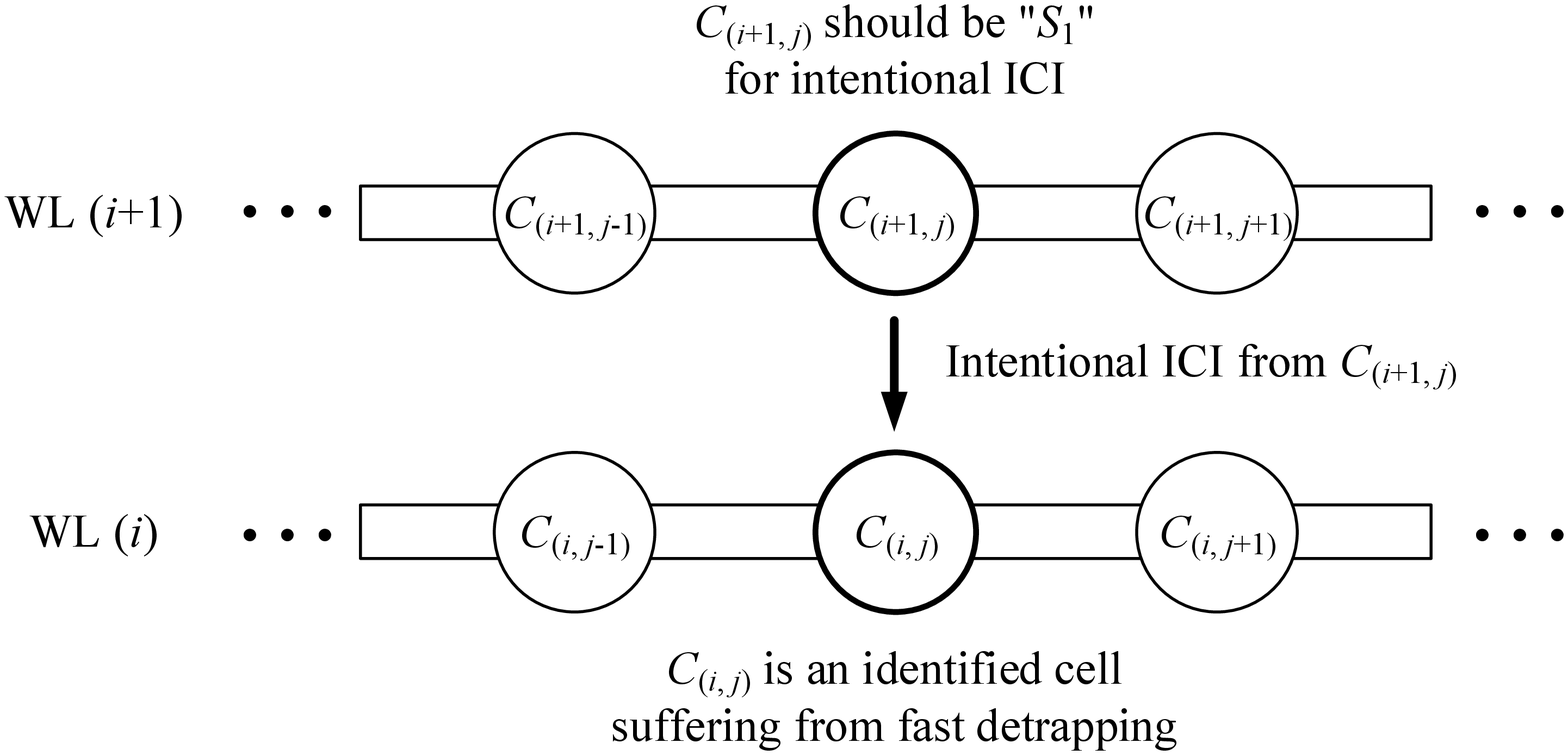}
   \caption{Intentional ICI for compensating fast detrapping. The cell $C_{(i+1, j)}$ will be regarded as stuck-at 0 (i.e., $S_1$) defect for the intentional ICI. (The index $k$ was omitted for simplicity.)}
   \label{fig:intentional_ici}
   \vspace{-2mm}
\end{figure}

\begin{figure}[t]
   \centering
   \includegraphics[width=0.29\textwidth]{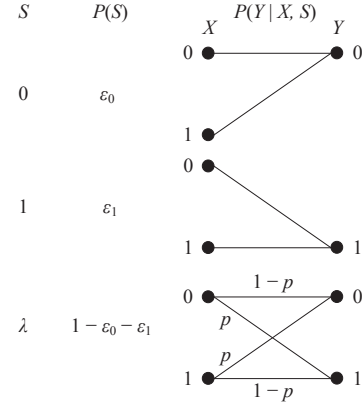}
   \caption{Channel model of a memory with defective cells.}
   \label{fig:defect_channel}
   \vspace{-5mm}
\end{figure}

The key idea is to compensate the effect of the fast detrapping in the $i$-th WL by controlling the \emph{intentional} ICI from the $(i+1)$-th WL. The intentional ICI can compensate the decrease of threshold voltage due to fast detrapping since the ICI increases the threshold voltage of the interfered cell. Note that fast detrapping results in charge loss, which decreases the corresponding cells' threshold voltages.

Assume that a cell $C_{(i, j)}$ in Fig.~\ref{fig:intentional_ici} suffered from fast detrapping, which has been identified by the read operations as explained in~\ref{subsec:know}. If the upper cell $C_{(i+1, j)}$ in the $(i+1)$-th WL is written into $S_1$, the ICI from $C_{(i+1, j)}$ will increase the threshold voltage of $C_{(i, j)}$ during increasing the threshold voltage of $C_{(i+1, j)}$ from $S_0$ to $S_1$. If $C_{(i+1, j)}$ is written into $S_0$, the threshold voltage of $C_{(i+1, j)}$ remains at the initial erase state $S_0$, so the ICI from $C_{(i+1, j)}$ is absent.

Thus, in order to compensate the effect of fast detrapping, the upper cell $C_{(i+1, j)}$ of the identified cell $C_{(i, j)}$ should be written into $S_1$ which is mapped into binary data ``0'' as shown in Fig.~\ref{fig:detrapping_before_after}~\subref{fig:detrapping_before}. We have to control the intentional ICI to alleviate the effect of fast detrapping.

We will formulate this problem of controlling the intentional ICI into \emph{coding in a memory with defective cells} in \cite{Kuznetsov1974}. The problem of coding in a memory with defective cells has been well studied in literature~\cite{Kuznetsov1974, Tsybakov1975additive, Heegard1983plbc}. A binary memory cell is called defective if its cell value is stuck-at a particular value regardless of the channel input. As shown in Fig.~\ref{fig:defect_channel}, this channel model has the ternary state $S \in \{0, 1, \lambda\}$, which will be the side information of defect. The state $S=0$ corresponds to a stuck-at 0 defect that always outputs a 0 independent of its input value, the state $S=1$ corresponds to a stuck-at 1 defect that always outputs a 1, and the state $S=\lambda$ corresponds to a normal cell that can be modelled by a binary symmetric channel (BSC) with crossover probability $p$. The probabilities of these states are $P(S=0) = \varepsilon_0$, $P(S=1) = \varepsilon_1$, and $P(S=\lambda) = 1 - \varepsilon_0 - \varepsilon_1$, respectively. Note that $X$ and $Y$ represent the channel input and output.

%If neither the encoder nor the decoder knows the side information of defect $S$, the capacity is given by
%\begin{equation} \label{eq:capacity_bdc_min}
%C_{\textrm{min}} = 1 - h \left(\left(1 - \varepsilon\right)p + \frac{\varepsilon}{2}\right)
%\end{equation}
%where $h\left( x\right) = -x \log_2 x - \left(1-x\right) \log_2\left(1-x\right)$. Note that \eqref{eq:capacity_bdc_min} equals the capacity of a BSC with crossover probability $\widetilde{p} = \left(1 - \varepsilon \right) p + \frac{\varepsilon}{2}$. If the encoder knows the side information of defect $S$, the maximum capacity of memory with defective cells can be achieved \cite{Heegard1983capacity}. The capacity is given by
%\begin{equation}\label{eq:capacity_bdc_max}
%C_{\max} = \left(1 - \varepsilon \right) \left( 1 - h \left(p \right) \right).
%\end{equation}

In order to solve the problem of coding in a memory with defective cells, Tsybakov proposed \emph{additive encoding} which masks defects by adding a carefully selected binary vector~\cite{Tsybakov1975additive}. Masking defects is to make a codeword whose values at the locations of defects match the stuck-at values at those locations~\cite{Tsybakov1975additive, Hwang2011a, Kim2013coding}. The additive encoding is a capacity achieving scheme~\cite{Heegard1983plbc, Kim2014duality}. %Heegard elaborated the additive encoding and defined the $[n, k, l]$ partitioned linear block codes (PLBC) that mask stuck-at defects and correct random errors \cite{Heegard1983plbc}.

Since the cell $C_{(i+1, j)}$ in Fig.~\ref{fig:intentional_ici} should store a binary 0 (i.e., $S_1$) for the intentional ICI, we will regard this cell $C_{(i+1, j)}$ as stuck-at 0 defects. Then, the additive encoding tries to make the codeword's element corresponding the cell $C_{(i+1, j)}$ become 0. After writing the additive encoded codeword into the $(i+1)$-th WL, the ICI from the cell $C_{(i+1, j)}$ increases the threshold voltage of the cell $C_{(i, j)}$, which compensates the threshold voltage decrease of the cell $C_{(i, j)}$ due to fast detrapping.

In summary, the encoder obtains the locations of identified cells which are suffering from fast detrapping in the $i$-th WL before writing the $(i+1)$-th WL. These locations can be obtained by one or two read operations, which would degrade the write speed performance. For the intentional ICI, the upper cells in the $(i+1)$-th WL of the identified cell in the $i$-th WL will be regarded as stuck-at 0 defects. Note that the side information of \emph{fast detrapping} in the $i$-th WL will be changed into the side information of \emph{defects} in the $(i+1)$-th WL. Afterwards, we can harness the intentional ICI by a coding technique such as additive encoding.

\section{Simulation Results}\label{sec:result}

\begin{table}[!t]
\renewcommand{\arraystretch}{1.3}
\caption{Simulation Parameters}
\label{tab:parameters}
\centering
{\small
\begin{tabular}{|c|c|}
\hline
Parameters & Values   \\ \hline \hline
Bits per cell & $B=1$ (SLC)  \\ \hline
%Architecture & All bitline (ABL) \\ \hline
Initial threshold voltage & \multirow{2}{*}{$\mathcal{N} \left(-4, 1^2\right)$}  \\
distribution & \\ \hline
Verify level for $S_1$ & $\nu_{S_1} = 1$  \\ \hline
Incremental step voltage & $\Delta V_\textrm{pp} = 1$ \\ \hline
Coupling ratio $\gamma$ of \eqref{eq:ici_simple} & 0.1 \\ \hline
$Z_\textrm{fast}$ of \eqref{eq:fmc2} & $\mathcal{N} \left(-0.2, \sigma_{Z_\textrm{fast}}^2 \right)$ \\ \hline
$Z_\textrm{random}$ of \eqref{eq:fmc2} & $\mathcal{N} \left(0, \sigma_{Z_\textrm{random}}^2 \right)$ \\ \hline
Read level & $\eta = 0$  \\ \hline
Identify level & $\zeta$  \\ \hline
\multirow{2}{*}{Additive encoding} & $\left[ n = 1023, k=923, l, r \right]$ \\
 & PBCH codes \\ \hline
\end{tabular}}
%\vspace{-3mm}
\end{table}

\begin{table}[t]
\renewcommand{\arraystretch}{1.3}
\caption{All Possible Redundancy Allocation Candidates of $\left[ n = 1023, k=923, l, r \right]$ PBCH Codes}
\label{tab:PLBC}
\centering
{\small
\begin{tabular}{|c|c|c|c|}
\hline
Code & {$l$} & {$r$} & Notes   \\ \hline \hline
0 & 0 & 100 & Only correcting random errors \\ \hline
1 & 10 & 90 &\\ \hline
2 & 20 & 80 &\\ \hline
3 & 30 & 70 &\\ \hline
4 & 40 & 60 &\\ \hline
5 & 50 & 50 &\\ \hline
6 & 60 & 40 &\\ \hline
7 & 70 & 30 &\\ \hline
8 & 80 & 20 &\\ \hline
9 & 90 & 10 &\\ \hline
10& 100 & 0 & Only masking defects\\ \hline
\end{tabular}}
\vspace{-5mm}
\end{table}

In this section, the simulation results are presented. The simulation parameters are summarized in Table.~\ref{tab:parameters}. The initial threshold voltage distribution (after erasing a flash memory block) is assumed to be the Gaussian distribution $\mathcal{N} \left(-4, 1^2\right)$. The ISPP was implemented with the parameters of the verify level for $S_1$, i.e., $\nu_{S_1} = 1$ and the incremental step voltage $\Delta V_\textrm{pp} = 1$. %Note that the variance of initial threshold voltage distribution and the incremental step voltage work for $Z_\textrm{write}$ of \eqref{eq:fmc1}, which precedes the ICI.

The noise due to fast detrapping $Z_\textrm{fast}$ is assumed to be the Gaussian distribution $\mathcal{N} \left(-0.2, \sigma_{Z_\textrm{fast}}^2 \right)$ by taking into account experimental results in~\cite{Chen2010}. The random noise due to other read noise sources $Z_\textrm{random}$ is assumed to the $\mathcal{N} \left(0, \sigma_{Z_\textrm{random}}^2 \right)$.

For the additive encoding, we use $\left[ n = 1023, k=923, l, r \right]$ partitioned Bose, Chaudhuri, Hocquenghem (PBCH) codes where $n$, $k$, $l$, and $r$ denote the codeword size, the information size, the redundancy size for masking defects, and the redundancy size for correcting random errors, respectively. The PBCH code is a special class of partitioned linear block codes, which can be designed by a similar method of standard BCH codes \cite{Heegard1983plbc}. For the given $n=1023$ and $k=923$ (i.e., the total redundancy size is 100), all possible redundancy allocation candidates of PBCH codes are presented in Table~\ref{tab:PLBC}~\cite{Kim2013redundancy}. Note that $l$ and $r$ are multiples of 10 that is the degree of Galois field. For the efficient computation complexity, two-step encoding scheme of \cite{Kim2013coding, Kim2013redundancy} has been used for encoding of PBCH codes.

\begin{figure}[!t]
   \centering
   \includegraphics[width=3in]{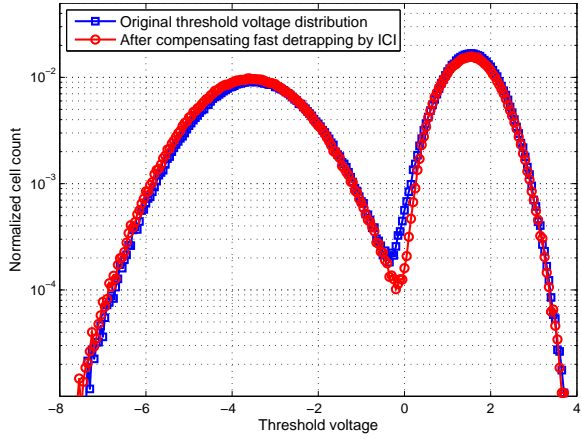}
   \caption{Comparison of threshold voltage distributions ($\sigma_{{Z_\textrm{fast}}} = 0.4$, $\sigma_{{Z_\textrm{random}}} = 0.2$, $\zeta = 0.4$).}
   \label{fig:distribution}
%  \vspace{-5mm}
\end{figure}

\begin{figure}[!t]
   \centering
   \includegraphics[width=3in]{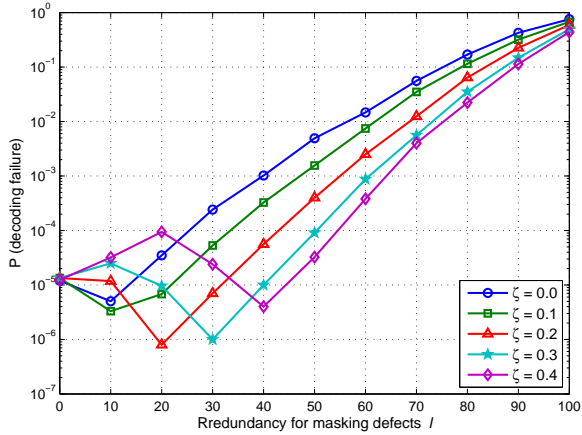}
   \caption{$P(\text{decoding failure})$ for different identify levels $\zeta$ ($\sigma_{{Z_\textrm{fast}}} = 0.4$, $\sigma_{{Z_\textrm{random}}} = 0.2$).}
   \label{fig:plot_wer_id}
  \vspace{-5mm}
\end{figure}

\begin{figure}[t]
\centering
\subfloat[Comparison of $P(\text{decoding failure})$ for different $\sigma_{Z_\textrm{fast}}$]{\includegraphics[width=3in]{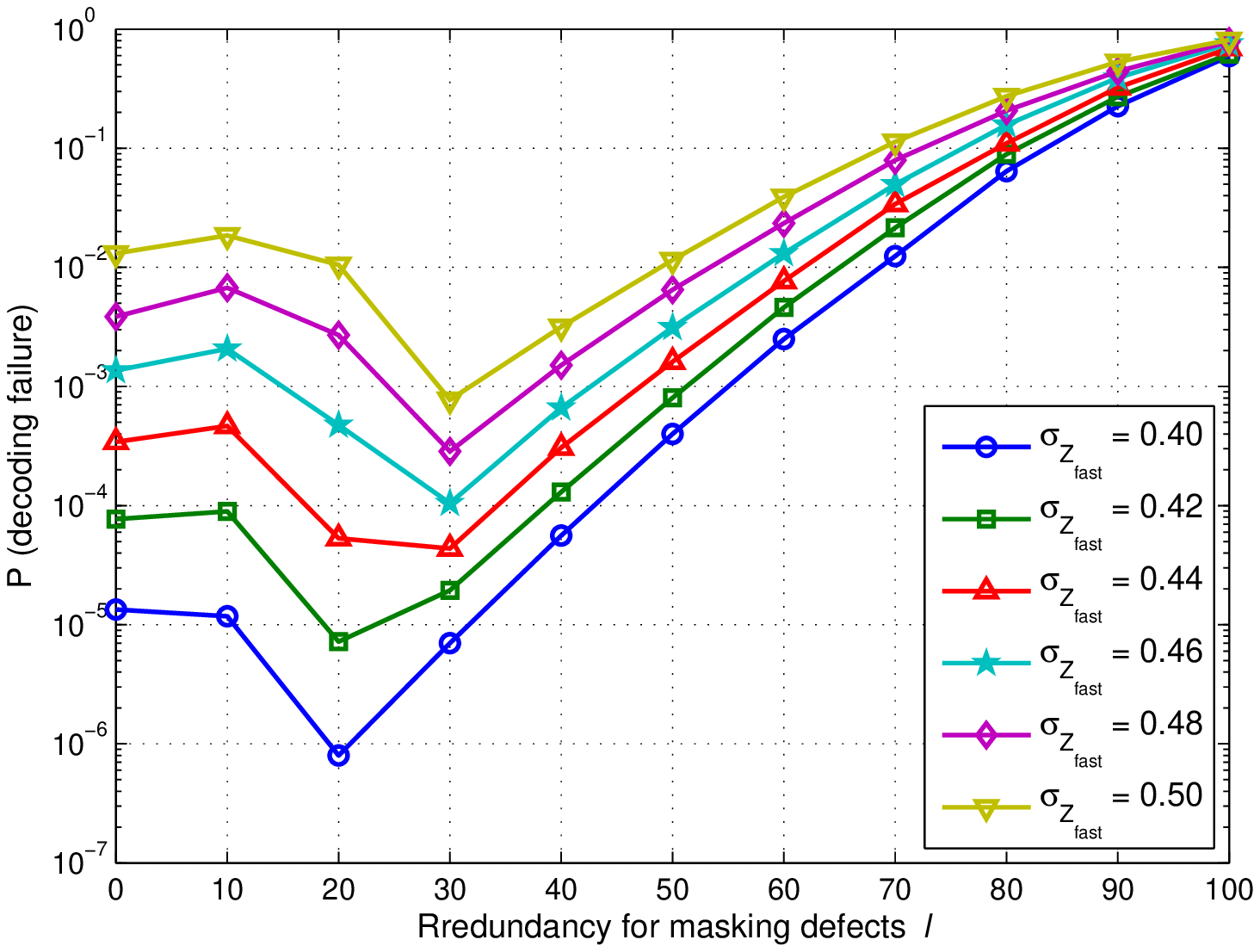}
\label{fig:plot_wer_fast_a}}
\hfil

\subfloat[Comparison of $P(\text{decoding failure})$ for $(l=0, r=100)$, (i.e., not using any side information) and that for the optimal redundancy allocation $\left(l^*, r^*\right)$]{\includegraphics[width=3in]{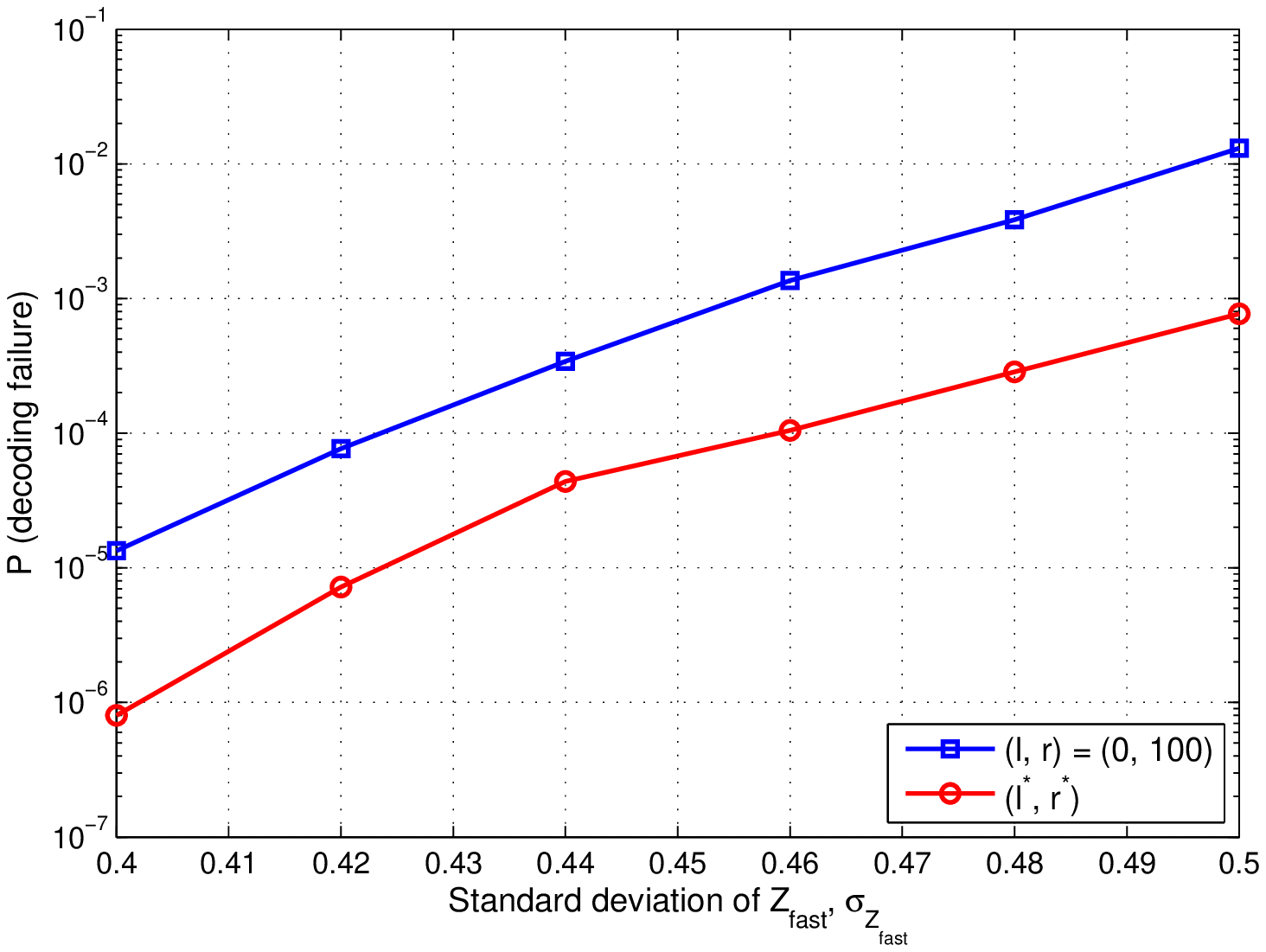}
\label{fig:plot_wer_fast_b}}

\caption{$P(\text{decoding failure})$ for different $\sigma_{Z_\textrm{fast}}$ ($\sigma_{{Z_\textrm{random}}} = 0.2$, $\zeta = 0.2$).}
\label{fig:plot_wer_fast}
\vspace{-5mm}
\end{figure}

\begin{figure}[t]
\centering
\subfloat[Comparison of $P(\text{decoding failure})$ for different $\sigma_{Z_\textrm{random}}$]{\includegraphics[width=3in]{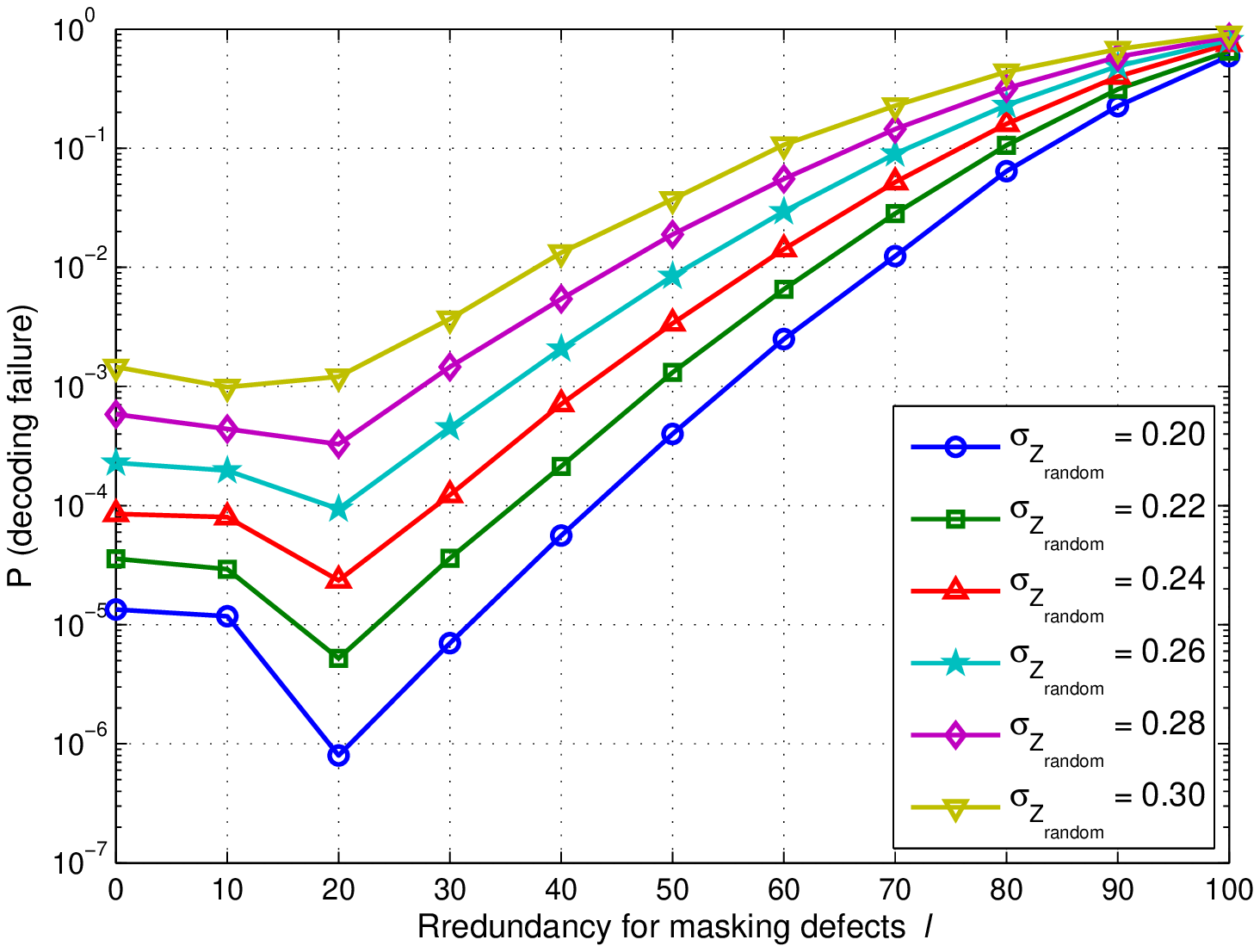}
\label{fig:plot_wer_random_a}}
\hfil

\subfloat[Comparison of $P(\text{decoding failure})$ for $(l=0, r=100)$, (i.e., not using any side information) and that for the optimal redundancy allocation $\left(l^*, r^*\right)$]{\includegraphics[width=3in]{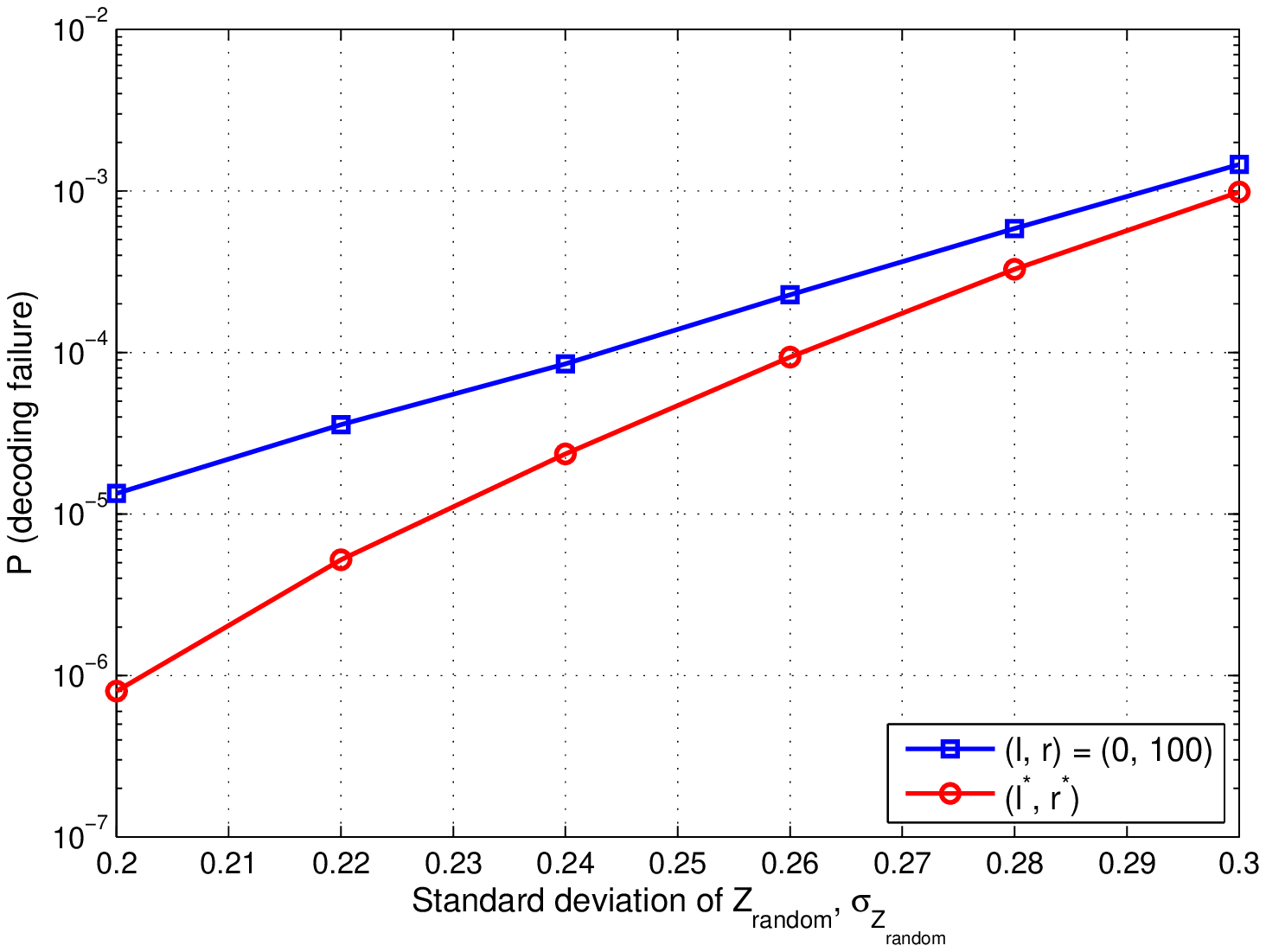}
\label{fig:plot_wer_random_b}}

\caption{$P(\text{decoding failure})$ for different $\sigma_{Z_\textrm{random}}$ ($\sigma_{{Z_\textrm{fast}}} = 0.4$, $\zeta = 0.2$).}
\label{fig:plot_wer_random}
\vspace{-5mm}
\end{figure}

Fig.~\ref{fig:distribution} shows that controlling ICI by additive encoding can compensate the effect of fast detrapping. After compensating the fast detrapping by the intentional ICI, the threshold voltage distribution improves.

Fig.~\ref{fig:plot_wer_id} shows that the probability of decoding failure $P(\textrm{decoding failure})$ is improved by the proposed scheme. If the redundancy for masking $l$ is zero, it means that the side information of fast detrapping is ignored. Otherwise, the encoder uses the side information of fast detrapping to improve $P(\textrm{decoding failure})$. Note that $P(\text{decoding failure})$ depends on the identify level $\zeta$ since it controls the number of identified cells, which is equivalent to the number of cells identified as defects in the upper WL. Also, the optimal redundancy allocation $\left(l^*, r^*\right)$ to minimize $P(\text{decoding failure})$ depends on the identify level $\zeta$. For larger $\zeta$, the number of cells identified as defects increases, which requires more redundancy for masking defects during additive encoding. For the given parameters in Fig.~\ref{fig:plot_wer_id}, the identify level of $\zeta = 0.2$ minimizes $P(\textrm{decoding failure})$.

Fig.~\ref{fig:plot_wer_fast} shows the improvement of $P(\textrm{decoding failure})$ by the proposed scheme for different levels of fast detrapping $Z_{\textrm{fast}}$. Note that $\sigma_{{Z_\textrm{fast}}}$ changes from 0.4 to 0.5 for the given $\sigma_{{Z_\textrm{random}}} = 0.2$. From Fig.~\ref{fig:plot_wer_fast}~\subref{fig:plot_wer_fast_a}, we can obtain the optimal redundancy allocation $\left(l^*, r^*\right)$ minimizing $P(\textrm{decoding failure})$. It is worth mentioning that the optimal redundancy for masking $l^*$ increases for larger $\sigma_{{Z_\textrm{fast}}}$. This is because we should allot more redundancy for masking defects (i.e., compensating the fast detrapping) as $\sigma_{{Z_\textrm{fast}}}$ increases.

Fig.~\ref{fig:plot_wer_fast}~\subref{fig:plot_wer_fast_b} compares $P(\textrm{decoding failure})$ for $(l, r) = (0, 100)$, (i.e., coding without the side information of fast detrapping) and that for $\left(l^*, r^*\right)$ (i.e., coding with the side information of fast detrapping). By using the side information of fast detrapping, we can significantly improve $P(\textrm{decoding failure})$ for different $\sigma_{{Z_\textrm{fast}}}$.

Fig.~\ref{fig:plot_wer_random} shows the improvement of $P(\textrm{decoding failure})$ by the proposed coding scheme for different random noise $Z_{\textrm{random}}$. Note that $\sigma_{{Z_\textrm{random}}}$ changes from 0.2 to 0.3 for the given $\sigma_{{Z_\textrm{fast}}} = 0.4$. From Fig.~\ref{fig:plot_wer_random}~\subref{fig:plot_wer_random_a}, we can obtain the optimal redundancy allocation $\left(l^*, r^*\right)$ minimizing $P(\textrm{decoding failure})$. Fig.~\ref{fig:plot_wer_random}~\subref{fig:plot_wer_random_b} compares $P(\textrm{decoding failure})$ for $(l, r) = (0, 100)$, (i.e., coding without the side information of fast detrapping) and that for $\left(l^*, r^*\right)$ (i.e., coding with the side information of fast detrapping). It is worth mentioning that the improvement of $P(\textrm{decoding failure})$ becomes significant as the fast detrapping $Z_{\textrm{fast}}$ dominates the random noise $Z_{\textrm{random}}$.

\section{Conclusion}\label{sec:conclusion}

We proposed a scheme to compensate the effect of fast detrapping by intentional ICI. The main idea comes from the observation that ICI increases the threshold voltage of a cell whereas fast detrapping decreases the threshold voltage of corresponding cell. Additive encoding can control the intentional ICI by using the side information of fast detrapping. Although this paper focused on SLC flash memory, the proposed scheme can be extended to MLC flash memories.

%% Appendix:
%% If needed a single appendix is created by
%\appendix
%% If several appendices are needed, then the command
%\appendices
%% in combination with further \section-commands can be used.

%\appendix

%%% Use \section* for acknowledgement
%\section*{Acknowledgment}
%
%The authors would like to thank various sponsors for supporting
%their research.

%% References:
%% We recommend the usage of BibTeX:
%%

\IEEEtriggeratref{5}
\bibliographystyle{IEEEtran}
\bibliography{IEEEabrv,fast_detrap}

\begin{thebibliography}{10}
\providecommand{\url}[1]{#1}
\csname url@samestyle\endcsname
\providecommand{\newblock}{\relax}
\providecommand{\bibinfo}[2]{#2}
\providecommand{\BIBentrySTDinterwordspacing}{\spaceskip=0pt\relax}
\providecommand{\BIBentryALTinterwordstretchfactor}{4}
\providecommand{\BIBentryALTinterwordspacing}{\spaceskip=\fontdimen2\font plus
\BIBentryALTinterwordstretchfactor\fontdimen3\font minus
  \fontdimen4\font\relax}
\providecommand{\BIBforeignlanguage}[2]{{%
\expandafter\ifx\csname l@#1\endcsname\relax
\typeout{** WARNING: IEEEtran.bst: No hyphenation pattern has been}%
\typeout{** loaded for the language `#1'. Using the pattern for}%
\typeout{** the default language instead.}%
\else
\language=\csname l@#1\endcsname
\fi
#2}}
\providecommand{\BIBdecl}{\relax}
\BIBdecl

\bibitem{Prall2007}
K.~Prall, ``{Scaling non-volatile memory below 30nm},'' in \emph{Proc. 22nd
  {IEEE} Non-Volatile Semiconductor Memory Workshop}, Aug. 2007, pp. 5--10.

\bibitem{Park2014scaling}
Y.~Park, J.~Lee, S.~S. Cho, G.~Jin, and E.~Jung, ``{Scaling and reliability of
  NAND flash devices},'' in \emph{Proc. IEEE Int. Reliab. Phys. Symp.}, Jun.
  2014, pp. 2E.1.1--2E.1.4.

\bibitem{Tanaka2007}
H.~Tanaka, M.~Kido, K.~Yahashi, M.~Oomura, R.~Katsumata, M.~Kito, Y.~Fukuzumi,
  M.~Sato, Y.~Nagata, Y.~Matsuoka, Y.~Iwata, H.~Aochi, and A.~Nitayama, ``{Bit
  cost scalable technology with punch and plug process for ultra high density
  flash memory},'' in \emph{Proc. IEEE Symp. VLSI Technol.}, Jun. 2007, pp.
  14--15.

\bibitem{Jang2009}
J.~Jang, H.-S. Kim, W.~Cho, H.~Cho, J.~Kim, S.~I. Shim, Y.~Jang, J.-H. Jeong,
  B.-K. Son, D.~W. Kim, K.~Kim, J.-J. Shim, J.~S. Lim, K.-H. Kim, S.~Y. Yi,
  J.-Y. Lim, D.~Chung, H.-C. Moon, S.~Hwang, J.-W. Lee, Y.-H. Son, U.-I. Chung,
  and W.-S. Lee, ``{Vertical cell array using TCAT (Terabit Cell Array
  Transistor) technology for ultra high density NAND flash memory},'' in
  \emph{Proc. IEEE VLSI Technol. Symp.}, Jun. 2009, pp. 192--193.

\bibitem{Park2014isscc}
K.-T. Park, J.-M. Han, D.~Kim, S.~Nam, K.~Choi, M.-S. Kim, P.~Kwak, D.~Lee,
  Y.-H. Choi, K.-M. Kang, M.-H. Choi, D.-H. Kwak, H.-W. Park, S.-W. Shim, H.-J.
  Yoon, D.~Kim, S.-W. Park, K.~Lee, K.~Ko, D.-K. Shim, Y.-L. Ahn, J.~Park,
  J.~Ryu, D.~Kim, K.~Yun, J.~Kwon, S.~Shin, D.~Youn, W.-T. Kim, T.~Kim, S.-J.
  Kim, S.~Seo, H.-G. Kim, D.-S. Byeon, H.-J. Yang, M.~Kim, M.-S. Kim, J.~Yeon,
  J.~Jang, H.-S. Kim, W.~Lee, D.~Song, S.~Lee, K.-H. Kyung, and J.-H. Choi,
  ``{Three-dimensional 128Gb MLC vertical NAND Flash-memory with 24-WL stacked
  layers and 50MB/s high-speed programming},'' in \emph{Proc. IEEE Int.
  Solid-State Circuits Conf. Dig. Tech. Pap. (ISSCC)}, Feb. 2014, pp. 334--335.

\bibitem{Chen2010}
C.-P. Chen, H.-T. Lue, C.-C. Hsieh, K.-P. Chang, and C.-Y. Lu, ``{Study of fast
  initial charge loss and it's impact on the programmed states Vt distribution
  of charge-trapping NAND Flash},'' in \emph{Proc. IEEE Int. Electron Devices
  Meet. (IEDM)}, Dec. 2010, pp. 5.6.1--5.6.4.

\bibitem{Lee2002}
J.-D. Lee, S.-H. Hur, and J.-D. Choi, ``{Effects of floating-gate interference
  on NAND flash memory cell operation},'' \emph{{IEEE} Electron Device Lett.},
  vol.~23, no.~5, pp. 264--266, May 2002.

\bibitem{Kuznetsov1974}
A.~V. Kuznetsov and B.~S. Tsybakov, ``{Coding in a memory with defective
  cells},'' \emph{Probl. Peredachi Inf.}, vol.~10, no.~2, pp. 52--60,
  Apr.--Jun. 1974.

\bibitem{ElGamal2011}
A.~{El Gamal} and Y.-H. Kim, \emph{{Network Information Theory}}.\hskip 1em
  plus 0.5em minus 0.4em\relax Cambridge, U.K.: Cambridge University Press,
  2011.

\bibitem{Tsybakov1975additive}
B.~S. Tsybakov, ``{Additive group codes for defect correction},'' \emph{Probl.
  Peredachi Inf.}, vol.~11, no.~1, pp. 111--113, Jan.--Mar. 1975.

\bibitem{Heegard1983plbc}
C.~Heegard, ``{Partitioned linear block codes for computer memory with
  ``stuck-at'' defects},'' \emph{{IEEE} Trans. Inf. Theory}, vol.~29, no.~6,
  pp. 831--842, Nov. 1983.

\bibitem{Jagmohan2010coding}
A.~Jagmohan, L.~A. Lastras-Montano, M.~M. Franceschini, M.~Sharma, and
  R.~Cheek, ``{Coding for multilevel heterogeneous memories},'' in \emph{Proc.
  {IEEE} Int. Conf. Commun. (ICC)}, Cape Town, South Africa, May 2010, pp.
  1--6.

\bibitem{Kurkoski2013}
B.~M. Kurkoski, ``{Rewriting flash memories and dirty-paper coding},'' in
  \emph{Proc. {IEEE} Int. Conf. Commun. (ICC)}, Budapest, Hungary, Jun. 2013,
  pp. 4353--4357.

\bibitem{Kim2014dirtyflash}
Y.~Kim and B.~V.~K. {Vijaya Kumar}, ``Writing on dirty flash memory,'' in
  \emph{Proc. 52nd Annu. Allerton Conf. Commun., Control, Comput.}, Monticello,
  IL, USA, Oct. 2014, pp. 513--520.

\bibitem{Hennessy2002}
J.~L. Hennessy and D.~A. Patterson, \emph{Computer Architecture: A Quantitative
  Approach}, 3rd~ed.\hskip 1em plus 0.5em minus 0.4em\relax San Francisco, CA,
  USA: Morgan Kaufmann Publishers Inc., 2002.

\bibitem{Suh1995}
K.-D. Suh, B.-H. Suh, Y.-H. Lim, J.-K. Kim, Y.-J. Choi, Y.-N. Koh, S.-S. Lee,
  S.-C. Kwon, B.-S. Choi, J.-S. Yum, J.-H. Choi, J.-R. Kim, and H.-K. Lim, ``{A
  3.3 V 32 Mb NAND flash memory with incremental step pulse programming
  scheme},'' \emph{{IEEE} J. Solid-State Circuits}, vol.~30, no.~11, pp.
  1149--1156, Nov. 1995.

\bibitem{Dong2011soft}
G.~Dong, N.~Xie, and T.~Zhang, ``{On the use of soft-decision error-correction
  codes in NAND flash memory},'' \emph{{IEEE} Trans. Circuits Syst. {I}},
  vol.~58, no.~2, pp. 429--439, Feb. 2011.

\bibitem{Kim2012verify}
Y.~Kim, J.~Kim, J.~J. Kong, B.~V.~K. {Vijaya Kumar}, and X.~Li, ``{Verify level
  control criteria for multi-level cell flash memories and their
  applications},'' \emph{EURASIP Journal on Advances in Signal Processing},
  vol. 2012, no.~1, pp. 1--13, 2012.

\bibitem{Kim2013modulation}
Y.~Kim, B.~V.~K. {Vijaya Kumar}, K.~L. Cho, H.~Son, J.~Kim, J.~J. Kong, and
  J.~Lee, ``{Modulation coding for flash memories},'' in \emph{Proc. {IEEE}
  Int. Conf. Comput., Netw. Commun. (ICNC)}, San Diego, CA, Jan. 2013, pp.
  961--967.

\bibitem{Moon2013}
J.~Moon, J.~No, S.~Lee, S.~Kim, S.~Choi, and Y.~Song, ``{Statistical
  characterization of noise and interference in NAND flash memory},''
  \emph{{IEEE} Trans. Circuits Syst. {I}}, vol.~60, no.~8, pp. 2153--2164, Aug.
  2013.

\bibitem{Hwang2011a}
E.~Hwang, B.~Narayanaswamy, R.~Negi, and B.~V.~K. {Vijaya Kumar}, ``{Iterative
  cross-entropy encoding for memory systems with stuck-at errors},'' in
  \emph{Proc. {IEEE} Global Commun. Conf. (GLOBECOM)}, Houston, TX, USA, Dec.
  2011, pp. 1--5.

\bibitem{Kim2013coding}
Y.~Kim and B.~V.~K. {Vijaya Kumar}, ``{Coding for memory with stuck-at
  defects},'' in \emph{Proc. {IEEE} Int. Conf. Commun. (ICC)}, Budapest,
  Hungary, Jun. 2013, pp. 4347--4352.

\bibitem{Kim2014duality}
------, ``{On the duality of erasures and defects},'' \emph{arXiv preprint
  arXiv:1403.1897}, vol. abs/1403.1, 2014.

\bibitem{Kim2013redundancy}
------, ``{Redundancy allocation of partitioned linear block codes},'' in
  \emph{Proc. {IEEE} Int. Symp. Inf. Theory (ISIT)}, Istanbul, Turkey, Jul.
  2013, pp. 2374--2378.

\end{thebibliography}

\end{document}